\documentclass{revtex4}
\usepackage{graphicx}
\usepackage[]{epsfig}

\newcommand{\beq}{\begin{equation}}
\newcommand{\eeq}{\end{equation}}
\newcommand{\beqd}{\begin{displaymath}}
\newcommand{\eeqd}{\end{displaymath}}
\newcommand{\beqa}{\begin{eqnarray}}
\newcommand{\eeqa}{\end{eqnarray}}
\newcommand{\non}{\nonumber}
\renewcommand{\a}{\alpha}

\newcommand{\s}{\sigma}
\newcommand{\arctanh}{{\rm arctanh}}
\newcommand{\comment}[1]{}

\newcommand{\Tr}{{\rm Tr}\,}
\newcommand{\tb}{\tilde{b}}

\begin{document}

\title{Large Deviations of the Free-Energy in Diluted Mean-Field Spin-Glass}

\author{Giorgio Parisi$^{1,2}$ and Tommaso Rizzo}

\affiliation{$^{1}$Dipartimento di Fisica, Universit\`a di Roma ``La Sapienza'', 
P.le Aldo Moro 2, 00185 Roma,  Italy
\\
$^{2}$Statistical Mechanics and Complexity Center (SMC) - INFM - CNR, Italy}

\begin{abstract}
Sample-to-sample free energy fluctuations in spin-glasses display a markedly different behaviour in finite-dimensional and fully-connected models, namely Gaussian vs. non-Gaussian.
Spin-glass models defined on various types of random graphs are in an intermediate situation between these two classes of models and we investigate whether the nature of their free-energy fluctuations is Gaussian or not.
It has been argued that Gaussian behaviour is present whenever the interactions are locally non-homogeneous, {\it i.e.} in most cases with the notable exception of models with fixed connectivity and random couplings $J_{ij}=\pm \tilde{J}$.
We confirm these expectation by means of various analytical results. In particular we unveil the connection between 
the spatial fluctuations of the populations of populations of fields defined at different sites of the lattice and the Gaussian nature of the free-energy fluctuations. On the contrary on locally homogeneous lattices the populations do not fluctuate over the sites and as a consequence the small-deviations of the free energy are non-Gaussian and scales as in the Sherrington-Kirkpatrick model. 
\end{abstract}

\maketitle

\section{Introduction}

The problem of sample-to-sample free energy fluctuations in spin-glasses has attracted large interest in recent years both from the theoretical and numerical point of view \cite{B1,BKM,PALA,B2,KKLJH,PAL,AM,DDF,AMY,ABMM,martin2,TAMAS,PR1,PR2,PR3}.

In general at fixed system size $N$ the free energy {\it per spin} of a given sample is a random variable with mean $f_{N}$ and variance $\sigma_N^2$.
In the thermodynamic limit $f_N$ approaches a definite value $f_{typ}$ while $\sigma_N$ approaches zero: the free energy is self-averaging and does not depend on the sample. 
It is expected that in the thermodynamic limit, the rescaled variable $x=(f-f_N)/\sigma_N$ has a limiting probability distribution that describes the {\it small deviations} of the free energy, {\it i.e.} those that occur with finite probability and whose scale decreases with $N$.
On the other hand large deviations of the free energy, {\it i.e.} those that remains finite in the large $N$ limit have a probability that vanishes exponentially with the size of the system and are described by a large deviation function \cite{CPSV,PR1,PR2,PR3}.   

In finite-dimensional models one expects the small deviations of the free energy to have a Gaussian distribution with variance proportional to the volume of the system {\it i.e.} $\sigma_N \propto N^{-1/2}$ \cite{MATH,AM,TAMAS}. In mean-field models however the standard arguments leading to this expectation  no longer hold and more exotic situations can be observed.
Indeed the scaling of $\sigma_N$ with $N$ in the Sherrington-Kirkpatrick(SK) model \cite{MPV} has been largely studied numerically in recent years \cite{B1,BKM,PALA,B2,KKLJH,PAL,ABMM} and there is growing consensus that $\sigma_N \propto N^{-5/6}$. This scaling had been conjectured early in \cite{CPSV} using  a large deviations result by Kondor \cite{Kon1} under the assumption that there is a matching between large and small deviations (for a recent discussion see \cite{PR2}). 
The large deviation result of Kondor had been questioned in \cite{AM,DDF}
but was later proved to be correct \cite{TALA}. The distribution of {\it negative} large deviations was computed down to zero temperature and excellent agreement with numerical results was found \cite{PR1,PR2}.  
Currently there is no analytical tool to compute directly $\sigma_N$ nevertheless the recent computation of {\it positive} large deviations \cite{PR3} adds further support to the $5/6$ scaling.
Furthermore the evaluation of the small-deviation distribution of the SK model is an open problem. We just know that it is not a Gaussian function but it has been argued that the behaviour of its tails can be deduced from the large deviations behaviour \cite{PR3}.

Mean-field spin-glass models defined on random graphs are considered as something intermediate between  fully-connected  models (notably the SK model) and finite-dimensional models.
Indeed they have a mean-field nature but are more realistic because every spin interacts with a finite number of neighbours. Therefore it is natural to ask what is the behaviour of the free energy fluctuations in these models. In particular we can ask if the small deviation distribution is Gaussian and if the large deviation function is similar to the one found in the SK model.
The ground state properties of spin-glass models defined on the Bethe Lattice have been studied intensively in recent times \cite{BKM,martin2,B3,B4,LPHJ}.
Our investigation was motivated by the findings of Ref. \cite{BKM}. 
The authors of Ref. \cite{BKM} observed that both the scaling of $\sigma_N$ and the shape of probability distribution of the ground state energy of a spin-glass defined on random graphs with fixed connectivity depends on the distribution of the couplings.
In particular on a random graph with fixed connectivity and Gaussian distributed couplings $J_{ij}$  it turned out that $\sigma_N \propto N^{-1/2}$ (as in finite dimensional models) and that the skewness of the small-deviation distribution tends to zero at large $N$, consistently with the assumption that it is a Gaussian.
However in the case of fixed connectivity and bimodal distribution of the couplings  ($J_{ij}=\pm \tilde{J}$) they observed a scaling of $\sigma_N$ definitively different from $N^{-1/2}$, (possible values being $N^{-3/4}$ or $N^{-4/5}$) . Furthermore it turned out that the skewness of the small deviation distribution instead of vanishing at large $N$ tends to go to a finite value consistent with that of the SK model.

In this paper we compute the large deviations of the models considered by \cite{BKM}, {\it i.e.} spin-glass models defined on graphs with fixed connectivity.
We consider the following functional \cite{CPSV,PR1,PR2}:
\beq
\Phi_N(n,\beta)=-{1\over \beta n N}\ln \overline{Z_{J}(\beta)^n}\ ,
\label{defphi}
\eeq
where different systems (or samples) are labeled by $J$, $Z_{J}(\beta)$ is the partition function of a given sample and the
bar denotes the average over different disordered samples.  
The above functional is the generating function of the cumulants of the sample-to-sample fluctuations of the free-energy.
In order to determine the moments of the small deviation distribution we should first compute the derivatives with respect to $n$ of $\Phi_N(n)$ at $n = 0$, and then take the limit $N \rightarrow \infty$.
This apparently simple step is a complex and open problem also in the SK model \cite{PR1,PR2,PR3}; instead the opposite case in which one takes first the $N \rightarrow \infty$ limit and then the $n \rightarrow 0$ limit is tractable, see \cite{PR2} for a complete discussion of this issue.
This amounts to compute:
\beq
\Phi(n,\beta)=\lim_{N \rightarrow \infty} \Phi_N(n,\beta)
\eeq
It is well known that the probability of large
deviations is related to the function $\Phi(n,\beta)$. Indeed
\beq
\exp( -\beta n N \Phi(n,\beta))=\overline{Z_{J}(\beta)^n}= \overline{\exp(-nN \beta f_{J}(\beta))}\ ,
\eeq
where $ f_{J}$ is the system-dependent free energy {\it per spin}. The region of positive $n$ corresponds to fluctuations where the free energy is smaller than the typical one and the region on negative $n$ corresponds to fluctuations where the free energy is larger than the typical one.

We define the large deviation function for the free energy, $L(f)$, (that we will call in the following  the sample complexity because it is related to the number of samples with free energy equal to $f$) as the logarithm divided by $N$ of the probability density of samples with free energy per spin $f$  in the thermodynamic limit:
\beq
L (f) \equiv \lim_{N\to\infty}{\log(P_{N}(f)) \over N }\ .
\eeq
For large $N$ the majority of the samples has free energy per spin equal to $f_{typ}$, and all other values have exponentially small probability. Consistently $L(f)$ is less or equal than zero, the equality holding $f=f_{typ}$, i.e. $L (f_{typ})=0$. For some values of $f$ it is possible that $L (f)=-\infty$,
meaning that the probability of large deviations goes to zero faster than exponentially with $N$.
 In the thermodynamic limit the  function $\Phi(n,\beta)$ defined in eq. (\ref{defphi}) yields the Legendre transform of $L(f)$ \cite{CPSV}, indeed we have:
\beq
-\beta n \Phi(n)=- \beta n f+ L(f)
\label{prima}
\eeq
where $f$ is determined by the condition:
\beq
\beta n={\partial L \over \partial f} 
\label{prima2}
\eeq
and equivalently we have:
\beq
 L(f)=\beta n f-\beta n \Phi(n)
\label{prima3}
\eeq
where $\beta n$ is determined by the condition:
\beq
f={\partial n \Phi \over \partial n} .
\label{prima4}
\eeq
Note that while numerical methods are best suited to study small deviations (precisely because they are typical) present day theoretical methods deal mainly with large deviations. This is because at the theoretical level large deviations requires essentially treating a replica field theory at the mean-field level, while small deviations requires the computation of the loop corrections. Furthermore the problem is made even more complicated by the fact that these corrections are singular \cite{PR3}.
Nevertheless it is usually assumed that information on small deviations can be extracted from large deviations \cite{CPSV,PR2}.
In particular if the expansion of $n\Phi(n)$ around $n=0$ reads $n\Phi(n)=n f_{typ}+c_2 n^2$ one expects that $\sigma_N \approx  (-2 c_2/\beta)^{1/2} N^{-1/2}$ and that the small deviation function is a Gaussian.
In the SK model instead $n \Phi(n)=n f_{typ}+c_6 n^6$ for positive $n$ and  $n \Phi(n)=n f_{typ}$ for negative $n$ \cite{DFM} and this has led to the $5/6$ scaling prediction for $\sigma_N$ \cite{CPSV}. Correspondingly the small deviation distribution is not expected to be a Gaussian, as confirmed by the numerics \cite{BKM}. In this case however the constant $c_6$ is not related to the sixth moment of the small deviation distribution but rather to its right tail \cite{PR2}.
Note that the fact that the function $\Phi(n)$ is constant for negative $n$ leads to $L(f)=-\infty$ for $f>f_{typ}$, indeed in this region the probability vanishes as $\exp[-O(N^2)]$ \cite{PR3}.
This interpretation scheme allows us to make contact between our results and those of \cite{BKM}. Indeed we find that in random graphs with fixed connectivity the presence of an $n^2$ term in the expansion of $n \Phi(n)$ depends on the nature of the coupling distribution. 
For a generic distribution ({\it e.g.} Gaussian) of the coupling the term is present and the small deviations are expected to be Gaussian. Nevertheless if the couplings have a bimodal distribution $J_{ij}=\pm \tilde{J}$ the $n^2$ term is absent and $\Phi(n)$ has the same qualitative properties of the SK model, namely $n \Phi(n)=n f_{typ}+c_6 n^6$ (with a different $c_6$) for positive $n$ and  $n \Phi(n)=n f_{typ}$ for negative $n$. Thus in the bimodal case we expect that $\sigma_N \propto N^{-5/6}$  and that the small deviation distribution is not a Gaussian, possibly the same of the SK model \cite{PR3}. With respect to the data of \cite{BKM} we think that the expected $5/6$ scaling is seen only at high enough system size as  in the case of the SK model \cite{ABMM}.

Our results on $\Phi(n)$ are thus in agreement with the findings of \cite{BKM}. Interestingly enough the authors of \cite{BKM}  suggested that the peculiar behavior of free energy fluctuations for bimodal distribution of the couplings is caused by the fact that the Hamiltonian is locally homogeneous.
Indeed around a given site the negative couplings can be transformed in positive coupling through a Gauge transformation.
The process must stop when the one of the sites already encountered in the process is reached again due to a loop of the graph,  however since loops are typically large the disordered nature of the model appears ``at infinity''.
As we will see in the following our results put this intuition on a firm ground. 
Indeed we will show that are precisely the spatial inhomogeneities of the interactions that generate a $n^2$ term in $n\Phi(n)$.
On the contrary in the bimodal case, the local homogeneity allows to obtain a solution that does not fluctuate over the sites (in a sense to be specified below) and this guarantees that $\Phi(n)=n f_{eq}$ for positive and negative $n$ both in the Replica-Symmetric (RS) case and for $n<0$ in the Replica-Symmetry-Breaking (RSB) case much as in the SK model.

\comment{
In brief the nature of  the potential $\Phi(n)$ of the spin-glass with random coupling constants $J_{ij}$ defined on diluted lattices can be summarized as follows.
The general criterion to determine the properties of $\Phi(n)$ is to look at the spatial fluctuations of the interactions.
As soon as the interactions have {\it local} spatial fluctuations the power series of $n \Phi(n)$ contains a term $O(n^2)$ yielding Gaussian fluctuations both in the high and low temperature phase.
On the contrary in the case of fixed connectivity with $J_{ij}=\pm \tilde{J}$ there are non-Gaussian fluctuations (as in the Sherrington-Kirkpatrick model).
Thus we expect have $O(N^{-1/2})$ Gaussian fluctuations of the free energy of the samples in the following models: 
\begin{itemize}
\item  fluctuating connectivity and fixed number of interactions (the case of variable number of interactions yields trivially Gaussian fluctuations) for general distribution $\rho(J)$ both in the high and low temperature phase. If $J=\pm \tilde{J}$ the fluctuations are $O(1)$ in the high temperature phase.
\item fixed connectivity and distribution $\rho(J)$ of the couplings with support on more than two opposite values, both in the low and high temperature phase. 
\end{itemize}
}

The plan of the paper is the following.
In section \ref{fixed} we will write down the variational expression of $\Phi(n)$ for the spin-glass on a random lattice with fixed connectivity  (the Bethe lattice).
In sections \ref{RSgeneral} and \ref{RSBgeneral} we will consider respectively the RS and RSB solution.
In both cases we will show that local inhomogeneities lead to the presence of a $O(n^2)$ term whose coefficient can be expressed in terms of the spatial fluctuations of the local fields.
On the contrary when the interactions have a bimodal distribution the resulting
local homogeneity  allows to obtain a solution that does not fluctuate over the sites and leads to the vanishing of the $n^2$ term. This also implies that $\Phi(n)=n f_{eq}$ for positive and negative $n$ in the RS case and for $n<0$ in the full-RSB case much as in the SK model.
In the latter case in order to compute the first non-trivial term for positive $n$ we resort to an expansion of $\Phi(n)$ in powers of the order parameter. This is presented in
 section \ref{expansion}. We will confirm that in the bimodal case the $O(n^2)$ term vanish  and that the first non-trivial term in the expansion of $\Phi(n)$ is $O(n^5)$ much as in the SK model.
At the end we will give our conclusions and discuss some interesting consequences of our results.

\section{The functional $\Phi(n)$ of the Bethe lattice spin-glass}
\label{fixed}

In this section we discuss the potential $\Phi(n)$ of the spin-glass defined on the Bethe lattice with fixed connectivity $M+1$.
Following \cite{GDD} we express $\Phi(n)$ as a  variational functional of the order parameter $\rho(\sigma)$ that is a function defined on $n$ Ising spins $\s$.
The variational expression of the free energy reads:
\beq
n \beta \Phi (n)=M \ln \Tr_{ \{ \s \} } \rho^{M+1}( \s  )-\frac{M+1}{2}\ln \int  \Tr_{ \{ \s \}} \Tr_{ \{ \tau \}} \rho^M( \s  ) \rho^M( \tau  )\langle  \exp \beta J \sum_{\a}\s_{\a}\tau_{\a} \rangle
\label{PHIdil}
\eeq
Where the square brackets mean average with respect to the distribution of $J$.
The above expression has to be extremized with respect to $\rho(\sigma)$. We note that it is invariant under a rescaling of $\rho(\sigma)$ so that we can choose any normalization for it.
If we normalize $\rho(\sigma)$ to the corresponding variational equation in terms of $\rho(\sigma)$ reads:
\beq
\rho(\s)= \frac{\left\langle\Tr_{\tau} \rho^M(\tau) \exp J \s \tau \right\rangle}{\langle \Tr_{\tau,\s} \rho^M(\tau) \exp \beta J \s \tau \rangle}.
\label{varDIL}
\eeq
where $\sigma\tau\equiv \sum_{\a}\s_\a \tau_\a$.

In the next two sections we discuss the RS and RSB ansatz of $\rho(\sigma)$ that are characterized respectively by fields and distributions of fields. 
In the RS case we will find that a crucial condition in order {\it not} to have $O(N^{-1/2})$ Gaussian fluctuations is that the fields do not fluctuate over the sites (which is possible in the low temperature phase only if the interactions are locally homogeneous).
In the RSB case this condition becomes a condition of homogeneity of the populations of fields, meaning that the populations do not fluctuate.
Note that at one-step RSB level this corresponds to the fact that we have to consider the so-called factorized solution and the fact that fluctuating solutions are actually obtained at this level \cite{MP1} confirms that the true solution is full-RSB. 

In the RS case the homogeneity condition guarantees that the same solution valid at $n=0$ can be used at $n$ different from zero yielding $\Phi(n)=n f_{eq}$ exactly. Much as in the SK model we expect that this statement holds in the RSB case only for negative $n$, because the full-RSB solution at positive $n$ cannot be the same at $n=0$ if we require that $x_{min} \geq n$ where $x_{min}$ is the first breaking point of the $q(x)$.  Indeed the expansion in the order parameter of section \ref{expansion} shows that the model is mapped in the SK model with different coefficients and an explicit computation shows  $\Phi(n)$ for positive $n$ has on $O(n^5)$ behaviour as SK.

\section{The Replica-Symmetric Solution}
\label{RSgeneral}

In this section we study the replica symmetric ansatz on $\rho(\sigma)$. 
We normalize $\rho(\s)$ to one, following \cite{MP1}.
In the RS case $\rho(\sigma)$ is function of $\sum_a \s_a$ and is parameterized by a  function $R(u)$ as:
\beq
\rho(\sigma) =\int du R(u) \frac{\exp \beta u \sum_{a} \s_a}{ (2 \cosh \beta u)^n}
\eeq
where $R(u)$ must satisfy $\int du R(u) =1$ because of the normalization of $\rho(\s)$.
Accordingly we have:
\beq
\ln \Tr_{ \{ \sigma \} } \rho^{M+1}(\sigma)=\ln \int  \left( \frac{2\cosh \beta \sum_{i}^{M+1}u_i}{\prod_i^{M+1}2\cosh \beta u_i} \right)^n \prod_{i=1}^{M+1}R(u_i)du_i
\label{RSPRIMO}
\eeq
We are interested in evaluating what is the dependence on $n$ of the previous quantity for fixed $R(u)$.
Expanding in powers of $n$ we get:

\beqa
\ln \Tr_{ \{ \sigma \} } \rho^{M+1}(\sigma) & = &n \left[  \langle\langle A \rangle\rangle   
  \right]+\frac{n^2}{2} \left[ (\langle\langle A^2\rangle\rangle-\langle\langle A\rangle\rangle^2) \right]+O(n^3)
\label{RSPRIMOEXP}
\eeqa
where we have used:
\beqa
\langle\langle A^p \rangle\rangle & \equiv & \int \prod_{i=1}^{M+1}R(u_i)du_i \left( \ln \frac{2\cosh \beta \sum_{i}^{M+1}u_i}{\prod_i^{M+1}2\cosh \beta u_i} \right)^p
\label{defA}
\eeqa
thus we see that i.f.f. $R(u)=\delta(u-u_0)$ ({\it i.e.} $R(u)$ is concentrated on some value $u_0$) there is no $O(n^2)$ term. On the other hand it is easily seen that in this case there are no higher terms as well, and the following relationship is valid at all orders in $n$: 
\beq
\ln \Tr_{ \{ \sigma \} } \rho^{M+1}(\sigma)  =  n  ( \ln 2\cosh \beta (M+1) u_0 -  (M+1)\ln 2\cosh \beta u_0 )
\eeq

So the crucial condition in order not to have a $O(n^2)$ term  is that $R(u)$ is a delta function, and a sufficient condition for this is that the interactions are locally homogeneous.  
The other term entering the free energy can be expressed as:
\begin{displaymath}
\ln \Tr_{\{\s\}} \Tr_{\{\tau\}} \rho^M(\s)\rho^M(\tau) \langle \exp (\beta \sum_{a}\s_a \tau_a) \rangle =
\end{displaymath}
\beq
= \ln \int \prod_{i=1}^{M}R(u_i)du_i  \prod_{i=1}^{M}R(v_i)dv_i \left\langle \left(  \sum_{\s,\tau} \frac{\exp[ \beta \sum_{i}^{M}u_i \s +\beta \sum_{i}^{M}v_i \tau+ \beta J \s \tau ]}{\prod_i^{M}4\cosh \beta u_i\cosh \beta v_i}  \right)^n  \right\rangle
\label{RSSECONDO}
\eeq
where the square brackets mean average with respect to the distribution of $J$.
As above we can expand in powers of $n$ and obtain:

\beqa
\ln \Tr_{\{\s\}} \Tr_{\{\tau\}} \rho^M(\s)\rho^M(\tau) \langle \exp (\beta \sum_{a}\s_a \tau_a) \rangle  & = & n \left[  \langle\langle B \rangle\rangle   
  \right]+\frac{n^2}{2} \left[ (\langle\langle B^2\rangle\rangle-\langle\langle B\rangle\rangle^2) \right]+O(n^3)
\label{RSSECONDOEXP}
\eeqa
where we have used:
\beqa
\langle\langle B^p \rangle\rangle & \equiv & \int \prod_{i=1}^{M}[R(u_i)du_iR(v_i)dv_i]   \left\langle
\left(  \ln \sum_{\s,\tau}\frac{\exp[ \beta \sum_{i}^{M}u_i \s +\beta \sum_{i}^{M}v_i \tau+ \beta J \s \tau ]}
{\prod_i^{M}4\cosh \beta u_i\cosh \beta v_i}\right)^p \right\rangle
\label{defB}
\eeqa
thus we see that the $O(n^2)$ is absent if $R(u_i)=\delta(u_i)$ {\it and} $\rho(J)=\pm \tilde{J}$, according to the criterion of local homogeneity of the interactions. On the other hand the $O(n^2)$ term is present also if $R(u_i)=\delta(u_i)$ but $\rho(J)$ is not bimodal, {\it e.g.} in the high temperature phase of the corresponding model.

\subsection{Fluctuations of the Free Energy in the Replica-Symmetric Solution}

The presence of a $O(n^2)$ term in the large deviation function leads  naturally to assume that the small deviations of the free energy are  Gaussian.
A straightforward computation shows that the variance of the small-deviation is proportional to the coefficient $c_2$ of the $O(n^2)$ in $n \Phi(n)$ according to:
\beq
\langle \Delta F^2 \rangle-\langle \Delta F \rangle^2= -\frac{2 c_2}{\beta}N 
\label{variance}
\eeq
{\it Around $n=0$ the coefficient $c_2$ can be computed noticing that since the $\Phi(n)$ is stationary with respect to $R(u)$ the derivative with respect to $n$  of $\Phi(n)$ is given by its partial derivative with respect to $n$}. 

Using the definition of $\Phi(n)$, eq. (\ref{PHIdil}), and the expansions eq. (\ref{RSPRIMOEXP}) and eq. (\ref{RSSECONDOEXP}) we get:

\beqa
\beta n \Phi(n) & = & n \left[ M \langle\langle A \rangle\rangle   
 -  \frac{M+1}{2}  \langle\langle B \rangle\rangle \right]+\frac{n^2}{2} \left[M (\langle\langle A^2\rangle\rangle-\langle\langle A\rangle\rangle^2) -  \frac{M+1}{2}(\langle\langle B^2\rangle\rangle-\langle\langle B\rangle\rangle^2)\right]+O(n^3)
\eeqa
where we have used definitions (\ref{defA}) and (\ref{defB}).
The previous expression has to be evaluated using the variational $R(u)$ obtained at $n=0$.

The RS solution with $R(u)=\delta(u)$ is correct above the critical temperature specified by the condition $\langle \tanh^2 \beta_c J\rangle=1/M$.
Therefore we conclude that above the critical temperature $\Phi(n)$ in general has a term $O(n^2)$ different from zero and its coefficient $c_2$ is given by:
\beq
c_2=-{M+1 \over 4}\left( \langle (\ln \cosh \beta J)^2\rangle -\langle \ln \cosh \beta J\rangle^2\right)
\eeq 
Clearly this coefficient vanishes in the case of a bimodal distribution.
Above the critical temperature  the solution is $R(u)=\delta(u)$ also for $n \neq 0$ and $\Phi(n)$ reads:
\beq
\Phi(n)=-{\ln 2 \over \beta}-{M+ 1 \over 2 \beta n}\ln \langle \cosh^n \beta J\rangle
\eeq
again we see that in the case of a bimodal distribution $\Phi(n)$ does not depend on $n$.
In the low temperature phase we know that the $\rho(\sigma)$ is no longer a constant.
The correct parameterization in the low-temperature spin-glass phase is full-RSB. In the following section we will describe the RSB ansatz and show that in general the expansion of $n\Phi(n)$ has a $O(n^2)$ term. Nevertheless we will see that in the case of a bimodal distribution the $O(n^2)$ term vanishes and that $\Phi(n)$ is constant for $n<0$ much as in the SK model. In section (\ref{expansion}) we will compute the first non-trivial term in $\Phi(n)$ for $n>0$ and show that it is $O(n^5)$ much as in the SK model.

\section{The Replica-Symmetry-Breaking Solution}
\label{RSBgeneral}

The replica-symmetry breaking (RSB) parametrization of $\rho(\sigma)$ in terms of distributions of fields was presented in \cite{MP1} and we refer to that paper for an explanation of the main ideas underlying it.
In particular we will work in the Replica framework rather than using the cavity method. The resulting equations are the same but while the former allow a quicker derivation the latter unveils the physical meaning of the populations and the appearance of free energy shifts.

We introduce the field $u$ that parameterizes a distribution over the values of an Ising spin $\s$ according to the formula $P(\s)=\exp(\beta u \s)/2 \cosh \beta u$. We define a probability distribution (population) $P^{(0)}(u)$ of such fields a $0$-distribution, 
correspondingly a $1$-distribution is a probability distribution on probability  distributions (population of populations) and so on.
In the following a $k$-distribution will be written as $P^{(k)}$, and it defines a measure $P^{(k)}dP^{(k-1)}$ over the space of $k-1$-distributions.

In order to parameterize the function $\rho(\sigma)$ with $K$ steps of RSB we need:
\begin{itemize}
\item  a $K$-distribution $P^{(K)}$;
\item $K$ integers  $1 \leq x_1,\dots , x_K \leq n $ (as usual for $n < 1$ they  become real and the inequalities change sign);
\end{itemize}
In the following we will consider the parameters $1 \leq x_1,\dots , x_K \leq n $ fixed and consider just the dependency on the distributions. 
The construction is iterative and requires a set of functions $\rho_{P^{(k)}}(\sigma)$ of $x_{k+1}$ spins with $k=1,\dots,K+1$ (we define $x_{k+1} \equiv n$ and $x_0\equiv 1$). The normalization of $\rho_{P^{(k)}}$ is crucial, we choose to normalize all of them to $1$.
We define $\rho_{P^{(k)}}(\sigma)$ starting from $\rho_{P^{(k-1)}}(\sigma)$, we first divide the $x_{k+1}$ spins in $x_{k+1}/x_{k}$ groups $\{\s_{\cal C}\}$ of $x_{k}$ spins labelled by an index ${\cal C}=1,\dots ,x_{k+1}/x_{k}$. Then we have:
\beq
\rho_{P^{(k)}}(\s )=\int P^{(k)}dP^{(k-1)}\prod_{{\cal C}=1}^{x_{k+1}/x_{k}}\rho_{P^{(k-1)}}(\{\s\}_{\cal C} )
\label{itdef}
\eeq
Thus $\rho(\s)\equiv \rho_{P^{(K)}}(\sigma)$ is defined iteratively starting from  the Replica-Symmetric case corresponding to $k=0$:
\beq
\rho_{P^{(0)}}(\s )=\int P^{(0)}(u)du \prod_{i=1}^{x_1}\frac{\exp \beta u \s_i}{2 \cosh \beta u}
\eeq
With the above definitions it is possible to express the variational free energy (\ref{PHIdil}) in terms of the populations of populations in the same way as we derived eq. (\ref{RSPRIMO}) and eq. (\ref{RSSECONDO}).
In order to do that we need to introduce two  functions: $\Delta F_1^{(k)}[P_1^{(k)},\dots,P_{M+1}^{(k)} ]$ is a function of $M+1$ $k$-populations  and $\Delta F_{12}^{(k)}[P_1^{(k)},\dots,P_{2M}^{(k)} , J]$ is a function of $2M$ $k$-populations and a coupling constant $J$.
Their definition is iterative, {\it i.e.} the function at level $k$ is defined in term of the function at level $k-1$:
\beq
\Delta F_1^{(k)}(P_1^{(k)},\dots,P_{M+1}^{(k)})\equiv-{1 \over \beta x_{k+1}}\ln \int \left[ \prod_{i=1}^{M+1}   P_i^{(k)}dP_i^{(k-1)}\right]  e^{-\beta x_{k+1}\Delta F_1^{(k-1)} (P_1^{(k-1)},\dots,P_{M+1}^{(k-1)}) }
\label{defDF1}
\eeq
\beq
\Delta F_{12}^{(k)}(P_1^{(k)},\dots,P_{2M}^{(k)},J)\equiv-{1 \over \beta x_{k+1}}\ln \int \left[ \prod_{i=1}^{2M}   P_i^{(k)}dP_i^{(k-1)}\right]  e^{-\beta x_{k+1}\Delta F_{12}^{(k-1)} (P_1^{(k-1)},\dots,P_{2M}^{(k-1)},J) }
\label{defDF12}
\eeq
The above definitions have to be supplemented with the definitions for $k=0$ that read:
\beq
\Delta F_1^{(0)}(P_1^{(0)},\dots,P_{M+1}^{(0)})\equiv-{1 \over \beta x_{1}}\ln \int \left[ \prod_{i=1}^{M+1}   P_i^{(0)}du_i\right]  \left( \frac{2\cosh \beta \sum_{i=1}^{M+1}u_i}{\prod_{i=1}^{M+1}2\cosh \beta u_i}\right)^{x_1}\ .
\label{defDF10}
\eeq

\beq
\Delta F_{12}^{(0)}(P_1^{(0)},\dots,P_{2M}^{(0)},J) \equiv -{1 \over \beta x_{1}}\ln \int \left[ \prod_{i=1}^{2M}   P_i^{(0)}du_i\right] \left( \frac{\sum_{\s ,\tau} \exp[\beta \sum_{i=1}^M u_i \s+\beta \sum_{i=M+1}^{2 M} u_i \tau +\beta J \s \tau ]}{\prod_{i=1}^{2 M}2 \cosh \beta u_i}  \right)^{x_1}\ .
\label{defDF120}
\eeq

\subsection{The Functional $\Phi(n)$ in the Replica-Symmetry-Breaking solution}

The variational free energy expressed in term of the $K$-population $P^{(K)}$ that parameterizes $\rho(\sigma)$ reads:
\beq 
n \beta \Phi(n)= n M \beta \Delta F_1^{(K)} (P^{(K)},\dots, P^{(K)})   + {M+1 \over 2}\ln \langle e^{-\beta n\Delta F_{12}^{(K)} (P^{(K)},\dots, P^{(K)},J) } \rangle
\label{fvar12}
\eeq
where the square brackets mean average over the coupling constant $J$ and we have used the functions defined above.
The proof that the above expression is equivalent to (\ref{PHIdil}) is not very complicated and we will not report it. We just mention that it can be obtained in an iterative way much as we will do in the appendix for the variational equation.
In order to determine the $K$-population that extremizes (\ref{fvar12}) we need to solve the corresponding variational equations obtained differentiating it with respect to $P^{(K)}$.
An equivalent way to obtain  $P^{(K)}$ is to consider the variational equation (\ref{varDIL}) and rewrite it in terms of $P^{(K)}$, we will show how to do this in the appendix. For practical purposes this second method is to be preferred because the corresponding equations can be solved by means of a population dynamics algorithm \cite{MP1}, however
in order to study the small-$n$ behaviour of $\Phi(n)$ is more useful to consider that variational expression (\ref{fvar12}).

In general the $K$-population that extremizes (\ref{fvar12}) depends on the value of $n$. 
In order to determine the expression of $\Phi(n)$ at small values of $n$ we can expand expression (\ref{fvar12}) at the second order in $n$ around $n=0$. This expression is variational in $P^{(K)}$ and therefore the total second derivative of $n \phi(n)$ with respect to $n$  is equal to the partial second derivative. However we must keep in mind that $\Delta F_1^{(K)} $ and $\Delta F_{12}^{(K)}$ both have an implicit dependence from $n$, therefore in order to derive with respect to $n$ we make this dependence explicit by rewriting expression (\ref{fvar12}) as:
\beqa
n \beta \Phi(n) & = & - M   \ln \int \left[ \prod_{i=1}^{M+1}   P^{(K)}dP_i^{(K-1)}\right]  e^{-\beta n \Delta F_{1}^{(K-1)} (P_1^{(K-1)},\dots,P_{M+1}^{(K-1)}) }  + 
\nonumber
\\
& + & {M+1 \over 2} \ln \int \left[ \prod_{i=1}^{2M}   P^{(K)}dP_i^{(K-1)}\right] \langle e^{-\beta n \Delta F_{12}^{(K-1)} (P_1^{(K-1)},\dots,P_{2M}^{(K-1)},J) }\rangle
\label{pip}
\eeqa  
Expanding in powers of $n$ we get:
\beqa
\beta n \Phi(n) & = & n \left[ M \langle\langle A_1 \rangle\rangle   
 -  \frac{M+1}{2}  \langle\langle A_{12} \rangle\rangle \right]-\frac{n^2}{2} \left[M (\langle\langle A_1^2\rangle\rangle-\langle\langle A_1\rangle\rangle^2) -  \frac{M+1}{2}(\langle\langle A_{12}^2\rangle\rangle-\langle\langle A_{12}\rangle\rangle^2)\right]+O(n^3)
\nonumber
\eeqa
where we have defined:
\beqa
\langle\langle A_1^p \rangle\rangle & \equiv & \int \prod_{i=1}^{M+1} P^{(K)}dP_i^{(K-1)}\left[\Delta F_{1}^{(K-1)} (P_1^{(K-1)},\dots,P_{M+1}^{(K-1)})  \right]^p
\nonumber
\\
\langle\langle A_{12}^p \rangle\rangle & \equiv &  \int \prod_{i=1}^{2M} [P^{(K)}dP_i^{(K-1)}] \left\langle \left[ \Delta F_{12}^{(K-1)} (P_1^{(K-1)},\dots,P_{2M}^{(K-1)},J)  \right]^p \right\rangle
\nonumber
\eeqa
Thus we conclude that the coefficient $c_2$ of the $O(n^2)$ term in the small-$n$ power series  of $n \Phi(n)$ is given  by the coefficient of the above $O(n^2)$ term evaluated with the population $P^{(K)}$ corresponding  to $n=0$. 
Therefore in general $c_2$ will be non-zero and we will expect Gaussian fluctuations with variance given by eq. (\ref{variance}).

We note that the physical interpretation of the $K$-RSB ansatz is that on a given sample ({\it i.e.} a graph with a given  disorder realization) the local fields on a given sites are described by a $K-1$-population and the population $P^{(K)}$ represents the distribution over the sites of these $K-1$-populations \cite{MP1}.
The functions $\Delta F^{(K-1)}_1$ and $\Delta F^{(K-1)}_{12}$ are interpreted as the  free-energy variations (shifts) that are observed in the process of adding respectively a spin and a bond to a given graph \cite{MP1}.
Thus the interpretation of the above equation is that {\it the Gaussian fluctuations of the free energy are determined by the local fluctuations of the free energy shifts}. 

\subsection{The factorized solution}

The results of the preceding subsection tell us that the $O(n^2)$ term will be absent only if there are no spatial fluctuations in the distribution of the fields.
This corresponds to the fact that $P^{(K)}$ is given by the so-called {\it factorized solution} $P^{(K)}=\delta(P^{(K-1)}-P_0^{(K-1)})$.
Indeed it is easy to check that if $P^{(K)}$ is factorized the term $(\langle\langle A_1^2\rangle\rangle-\langle\langle A_1\rangle\rangle^2)$  vanishes. In order to have that also the term $(\langle\langle A_{12}^2\rangle\rangle-\langle\langle A_{12}\rangle\rangle^2)$ vanishes we need also the condition $J=\pm \tilde{J}$ (in the high-temperature phase the $O(n^2)$ term is determined solely by the fluctuations of $J$).

The variational equation for $P^{(K)}$ reported in the appendix shows that $P^{(K)}$ can be factorized only if the couplings have a bimodal distribution $J =\pm \tilde{J}$.
We argue that in this case the correct distribution is indeed factorized.
Note that for $K=1$ a non-factorized solution was found in \cite{MP1} for the bimodal case. We believe that this is an artifact of the fact that the correct solution has an infinite number of RSB steps $K=\infty$. 
This is similar to what happens for the function $q(x)$ of the SK model: in the 1RSB ansatz we find $q(0) \neq 0$ while $q(0)=0$ in the full-RSB solution \cite{MPV}.

In the bimodal case one can see that the factorized solution $P^{(K)}=\delta(P^{(K-1)}-P_0^{(K-1)})$ is such that   $P_0^{(K-1)}$ is independent of $n$. As a consequence $\Phi(n)$ is constant in $n$. Much as in the SK model we argue that $\Phi(n)$ is constant only for $n<0$, while for $n>0$ we have to abandon the factorized solution corresponding to $n=0$ because of the condition that the smallest RSB parameter $x_{min}$ must be larger than $n$.
In order to study this effect and to determine the first non-trivial term of $\Phi(n)$ for positive $n$ in the following section we will study an expansion of $\Phi(n)$ in the order parameter.

\section{Expansion of $\Phi(n)$ in powers of the order parameter}
\label{expansion}

In this section we report the expansion of  the potential $\Phi(n)$ of the spin-glass defined on the Bethe lattice with fixed connectivity $M+1$ in powers of the order parameter.
Note that following \cite{GDD} we will use a different normalization of $\rho(\s)$ with respect to the previous subsection, we write it as:
\beq
\rho(\{\s\})=\sum_{k=0}^n b_k \sum_{(\a_1 \dots \a_k)}q_{\a_1\dots\a_k}\s_{\a_1}\dots \s_{\a_k}
\eeq
with 
\beq
b_k \equiv \langle \cosh^n \beta J  \tanh^k \beta J\rangle \ \ \  {\rm and} \ \ \ \tilde{b}_k \equiv b_k/b_0 
\label{defb}
\eeq
The variational equation reads:
\beq
\rho(\{\s\})=\frac{\Tr_{ \{ \tau \}} \rho^M( \tau  )\langle  \exp \beta J \sum_{\a}\s_{\a}\tau_{\a} \rangle}{\Tr_{ \{ \tau \}} \rho^M( \tau  )}
\eeq
expressed in terms of the $q_{\a_1 \dots \a_k}$ reads:
\beq
q_{\a_1\dots \a_k}=\frac{\Tr_{ \{ \s \}} \s_{\a_1}\dots \s_{\a_k}\rho^M( \s )}{\Tr_{ \{ \s \}} \rho^M( \s  )}
\label{eqq}
\eeq
We have expanded expression (\ref{PHIdil}) in powers of the order parameter $q_{ab}$ at fourth order. The four-indexes order-parameter $q_{abcd}$ has been expressed in terms of $q_{ab}$ by means of its variational equation.
In the appendix we give some details while here we report the results:
\beq
n \beta \Phi =-\frac{M+1}{2}\ln b_0-n \ln 2+ F_{var}.
\label{phipow}
\eeq
Where 
\beq
F_{var} \equiv -{\tau\over 2}\Tr q^2-{\omega \over 6} \Tr q^3 -{v \over 8} \Tr q^4 + {y \over 4} \sum_{abc}q^2_{ab}q^2_{ac}-{u \over 12}\sum_{ab}q_{ab}^4+{\tilde{y} \over 4}(\Tr q^2)^2+O(q^5)
\label{fvar}
\eeq
where the various coefficients depends on $b_k$ with $k=2,4$, see their explicit expressions below.

In the high temperature region we have $q=0$ and therefore $F_{var}=0$, thus the only relevant term is the first term in eq. \ref{phipow}.
 If $J$ can take just two values $J=\pm \tilde{J}$ we have  $b_0=\cosh^n \beta \tilde{J}$, therefore for $T>T_c$ $\Phi(n)$ is just a constant in $n$. From a high temperature expansion we can verify that in this case the fluctuations of the extensive free energy scale as $\Delta F=O(1)$. On the contrary if the distribution of $J$ is not just peaked at $\pm \tilde{J}$,  $\Phi(n)$ is linear in $n$ and the fluctuations of the extensive free energy are Gaussian with $\Delta F=O(\sqrt{N})$. 
Therefore if the graph is locally homogeneous the first term immediately yields small $O(1)$ fluctuations above the critical temperature.

Below the critical temperature we have to check whether $n \Phi(n)$ is quadratic in $n$ around $n=0$.
We note that the various coefficients $\{ \tau, \omega , u, v, y\}$ in eq. (\ref{fvar}) depend on $n$ and on the temperature, however we first study the model with fixed coefficients and variable number of replicas $n$; we also reabsorb the coefficient $\tilde{y}$ in $y$, which is possible if we restrict the choice of the matrix $q_{ab}$ to those that verify the condition that $(q^2)_{aa}$ does not depend on $a$, this produce an additional dependence of $y$ from $n$ because $y+n \, \tilde{y} \rightarrow y$ but, like the other coefficients, we first consider it as independent of $n$.

We have computed the free energy of this model, basically generalizing Kondor's original computation to general values of the coefficients and including all the fourth-order terms.
We have found that much as in the SK model: i)  the first non linear term in the expansion of $n\,\Phi(n)$ for $n \geq 0$ is $O(n^6)$, due to non-trivial cancellations at order $O(n^2)$ and $O(n^4)$ and ii) $n\,\Phi(n)$ is just linear in $n$ for $n<0$. More explicitly we have for $n>0$
\beqa
F_{var} & = & n\left({\tau^3\over 6 \omega^2}+{2 u+ 9 v+6 y \over 24 \omega^4}\tau^4+O(\tau^5)\right)+
\nonumber
\\
& - & {9 \over 5120} n^6 \left( 40{\omega^7 \over u^6 }-75 {\omega^8 \over u^7}+36 { \omega^9 \over  u^8} +O(\tau)\right)+O(n^7) \ \ \ {\rm for} \ n>0
\label{fvarn}
\eeqa
In the SK model we have $\omega=u=v=y=1$ and we recover Kondor's value $-9/5120$ for the $O(n^6)$ coefficient.
Now to study the actual $n$-dependence of the model we have to take into account the fact that the coefficients $\{ \tau, \omega , u, v, y\}$ depend on $n$, indeed the various coefficient reads:
\beqa
{\tau} & = &  {1 \over 2} M\,\left( 1 + M \right) \,{{\tilde{b}_2}}^2\,
      \left( -1 + M\,{\tilde{b}_2} \right)   \, .
\label{tau}
\non
\\
 {\omega }& = &  M\,\left( -1 + M^2 \right) \,{{\tilde{b}_2}}^3\,
      \left( -2 + 3\,M\,{\tilde{b}_2} \right)   \, .
\non
\\
{v } & = &  \frac{M\,\left( -1 + M^2 \right) \,{{\tilde{b}_2}}^4\,
      \left( 3\,\left( -2 + M \right)  + M\,\left( -1 + 4\,M \right) \,{\tilde{b}_4} + 
        2\,M\,{\tilde{b}_2}\,\left( 5 - 3\,M + 
           \left( -3 + M \right) \,M\,{\tilde{b}_4} \right)  \right) }{-1 + 
      \,M\,{\tilde{b}_4}} \, .
\non
\\
{y } & = & - \frac{M\,\left( -1 + M^2 \right) \,{{\tilde{b}_2}}^4\,
      \left( 6 - 3\,M + \left( 8 - 11\,M \right) \,M\,{\tilde{b}_4} + 
        M\,{\tilde{b}_2}\,\left( -9 + 5\,M + 
           M\,\left( 1 + 3\,M \right) \,{\tilde{b}_4} \right)  \right) }{-1 + 
      \,M\,{\tilde{b}_4}} \, .
\non
\\
{u } & = & -\frac{M\,\left( -1 + M^2 \right) \,{{\tilde{b}_2}}^4\,
    \left( -6\,\left( -2 + M \right)  + 3\,\left( 3 - 5\,M \right) \,M\,{\tilde{b}_4} + 
      4\,M\,{\tilde{b}_2}\,\left( -2 + M + 
         M\,\left( -1 + 2\,M \right) \,{\tilde{b}_4} \right)  \right) }{\,
    \left( -1 + M\,{\tilde{b}_4} \right) } \, .
\non
\\
\tilde{y} & = & \frac{M\,\left( 1 + M \right) \,{{\tilde{b}_2}}^4\,
    \left( -3 + 4\,M + \left( 2 - 3\,M \right) \,M^2\,{\tilde{b}_4} + 
      M^3\,{{\tilde{b}_2}}^2\,\left(  M\,{\tilde{b}_4}-1 \right)  + 
      4\,\left(  M -1\right) \,M\,{\tilde{b}_2}\,
       \left(  M^2\,{\tilde{b}_4} -1 \right)  \right) }{4\,
    \left(  M\,{\tilde{b}_4} -1\right) } \, .
\label{yt}
\eeqa
In order to recover the SK limit of the above coefficients, we have to rescale the couplings as $J=\tilde{J}/M^{1/2}$ where $\tilde{J}$ is a random variable with unit variance and take the limit $M \rightarrow \infty$. We obtain $\tau=1-T+O(1-T)^2$ and $\omega=u=v=y=1$, while $\tilde{y}=0$ {\it i.e.} as it should.

Each coefficient depends on $\beta$ and $n$ through $\tilde{b}_k$ and we have:
\beq 
\tilde{b}_k=\langle \tanh^k \beta J\rangle+n \Big(\langle \ln \cosh \beta J \tanh^k \beta J \rangle -\langle \ln \cosh \beta J\rangle \langle \tanh^k \beta J \rangle\Big)+O(n^2)
\eeq
Therefore the coefficients for a generic distribution of the $J's$ have a linear dependence on $n$ that, included in expression (\ref{fvarn}), leads to a $O(n^2)$ dependence of $n \Phi(n)$.
Note that the coefficients  have a dependence from $n$ that is regular around $n=0$ and therefore the first term in eq. (\ref{fvarn}) is regular in $n$ around $n=0$, but there is still the $O(n^6)$ term which produces a non regular dependence from $n$ around $n=0$. In other words  in the general diluted model the function $n \phi(n)$ develops a regular $O(n^2)$ dependence but there is still a singularity at $n=0$ in the sixth derivative.
Near the critical temperature the leading $O(n^2)$ term is given by the $n$ dependence of $\tau$ in eq. (\ref{fvarn}) and therefore is $O(\tau^2)$, expanding eq. (\ref{tau}) in powers of $n$ we get for the $O(n^2)$ term of $F_{var}$: 
\beq
n^2 \left({M^4 \over 4(M-1)^2(M+1)} \tau_0^2 \Big(\langle \ln \cosh \beta_c J \tanh^2 \beta_c J \rangle -\langle \ln \cosh \beta_c J\rangle \langle \tanh^2 \beta_c J \rangle\Big) +O(\tau_0^3)\right)
\eeq 
Where  $\tau_0$ is given by eq. (\ref{tau}) computed at $n=0$. 

We consider now the locally homogeneous case in which $J =\pm \tilde{J}$ with equal probability. We note first that in this case the coefficients $\tilde{b}_k$  do not depend on $n$ anymore (see eq. \ref{defb}) and as a consequence $\tau, \omega, u $ and $v$ do not depend on $n$ neither. Thus the only dangerous coefficient is $y$ in which we have reabsorbed the coefficient $\tilde{y}$ through $y +n\, \tilde{y} \rightarrow y$.
The $\tilde{y}$ coefficient turns out to be zero
 and therefore  the behavior of the model is the same of the SK model, the first non linear term in $n \phi(n)$ being $O(n^6)$.
To be more precise we have checked that there are no $O(n^2)$ terms in $n \phi(n)$ at the first non-trivial order in $\tau$, because $\tilde{y}$ defined according to eq. (\ref{yt}) actually is zero only at the critical temperature where $\tilde{b}_2=1/M$ and $\tilde{b}_4=1/M^2$ (because $J =\pm \tilde{J}$) but has small non-zero corrections $O(\tau)$ at higher orders. These $O(n^2)$ corrections are likely cancelled by the $O(n^2)$ higher order term $\Tr Q^2 \Tr Q^3$ not included in the computation. 

In the appendices we report a similar expansion for the variational equation and various quantities relevant for the computation.

\section{Conclusion}
We have investigated the large deviations free energy functional $\Phi(n)$ in the case of the Bethe lattice spin-glass and we have confirmed that Gaussian behaviour of the free energy fluctuations has to be expected whenever there is no local homogeneity of the interactions. In particular the only case in which we found a non-Gaussian SK-like behaviour in when the random couplings can take only two possible opposite values $J_{ij}=\pm \tilde{J}$ with equal probability.

In general the quantity $\Phi(n)$ can be expressed in terms of a distribution of fields. In the RS case we have a single distribution corresponding to the possible values of the cavity fields at different sites of the lattice for a given disorder realization. In the RSB phase we have a population of populations, {\it i.e.} on each site we have a population of fields corresponding to the presence of many states.
We have found that if the system is locally homogeneous we can find a locally homogenates distribution of the fields and this leads to the vanishing of the $O(n^2)$ term in $\Phi(n)$.
Thus we argue that the correct RSB solution in the bimodal case is the so-called factorized solution.
Instead if the system is not locally homogenates the $O(n^2)$ terms in $\Phi(n)$ can be evaluated using the $n=0$ solution because of stationarity.

We have also verified that the expansion in power of the order parameter near the critical temperature in the locally homogenates case is formally equivalent to that of the SK model and found that $\Phi(n)$ has the same $O(n^5)$ behaviour of SK for small positive $n$.

We note that the fact that in the bimodal case $n \Phi(n)=n f_{typ}$ for $n<0$ has some interesting consequences.
Indeed since $\Phi(n)$ is the Legendre transform of the large deviations function $L(f)$ (see eqs. (\ref{prima},\ref{prima2},\ref{prima3},\ref{prima4})) it follows that $L(f)=-\infty$ for free energies {\it per spin} larger than the typical one $f_{typ}$. This means that the probability of finding a sample with $f>f_{typ}$ is smaller than $\exp[O(N)]$. Indeed for the SK model a recent computation \cite{PR3} has shown that $P(f) \propto \exp[O(N^2)]$.
This scaling cannot hold for the Bethe lattice because while the total number of samples is actually $\exp[O(N^2)]$ in the SK model, the total number of samples on the Bethe lattice is $\exp[(M+1)N \ln N]$ at leading order.
Thus we argue that in the Bethe lattice with bimodal distribution of the couplings $P(f) \propto \exp[O(N\ln N)]$ for $f>f_{typ}$ although the actual computation is beyond the scope of this work. For $M=1$ detailed computations are easy.
Nevertheless we note that free energies larger than the typical one can only be observed on graphs with topologies different from the typical one, ({\it e.g.} a regular lattice). In other words { \it the probability of observing a free energy (and in particular a ground state energy) larger than the typical one on a graph with typical topology is strictly zero}.
Indeed suppose that by just changing the signs of the interactions of a typical graph ({\it i.e.} without modifying the incidence matrix) we could raise the free energy {\it per spin}.  
Since the number of links on a graph is precisely ${M+1 \over 2}N$ the probability of such a sample will be $\exp[O(N)]$ and this would lead to a non-constant $\Phi(n)$ for $n<0$ contrarily to what we have computed.

\comment{

\section{The Spin-Glass on the Poissonian Random Graph}

For the random graph with poissonian distribution of the connectivities we have the following expression for the the functional $\Phi(n)$:
\beq
n \beta \Phi(n)=-{\gamma  \over 2} + {\gamma \over 2}\Tr_{\{\s,\tau\}}\rho(\s)\rho(\tau)\langle \exp [-\beta J \sum_a\s_a \tau_a ]\rangle-\Tr_{\{\s\}}\rho(\s)\ln \rho(\s) 
\eeq

}

\appendix

\section{The Variational Equations in terms of Populations}

In this appendix we write the variational equations in terms of populations.
These equations have been obtained at the level of one-step RSB in \cite{MP1} using the cavity method. In the following we write them down for a generic number of RSB steps using the replica method.
The variational equation that extremizes the free energy  (\ref{PHIdil}) reads:
\beq
\rho(\s)= \frac{\left\langle\Tr_{\tau} \rho^M(\tau) \exp J \s \tau \right\rangle}{\langle \Tr_{\tau,\s} \rho^M(\tau) \exp \beta J \s \tau \rangle}.
\label{varr}
\eeq
 in terms of the $K$-population the above equation reads:
\beqa
P^{(K)} & \equiv & {1 \over \langle e^{-\beta n\Delta F^{(K)} (P^{(K)},\dots, P^{(K)},J) } \rangle}\int \left[ \prod_{i=1}^M   P^{(K)}dP_i^{(K-1)}\right] 
\times
\non
\\
& \times & \langle \delta(P^{(K-1)}-\tilde{P}^{(K-1)}) e^{-\beta n\Delta F^{(K-1)} (P_1^{(K-1)},\dots,P_M^{(K-1)},J) } \rangle
\label{varP}
\eeqa
Where the square brackets mean average over the disorder. 
In the above equation we have used the following functions of populations:
 i) a function $\tilde{P}^{(k)}[P_1^{(k)},\dots,P_M^{(k)} , J]$ that yields a $k$-population from $M$ other $k$-populations and ii) a function $\Delta F^{(k)}[P_1^{(k)},\dots,P_M^{(k)} , J]$ (also called the free-energy shift \cite{MP1}) that yields a real number from $M$ $k$-populations. The definition is iterative: the function $\tilde{P}^{(k)}$ and $\Delta F^{(k)}$ at level $k$ of RSB are defined starting from the functions $\tilde{P}^{(k-1)}$ and $\Delta F^{(k-1)}$:
\beqa
\tilde{P}^{(k)}(P_1^{(k)},\dots,P_M^{(k)},J) & \equiv & {1 \over e^{-\beta x_{k+1}\Delta F^{(k)} (P_1^{(k)},\dots,P_M^{(k)},J) }}\int \left[ \prod_{i=1}^M   P_i^{(k)}dP_i^{(k-1)}\right] 
\delta(P^{(k-1)}-\tilde{P}^{(k-1)})
\times
\non
\\
& \times & e^{-\beta x_{k+1}\Delta F^{(k-1)} (P_1^{(k-1)},\dots,P_M^{(k-1)},J) }
\label{defPnuApp}
\eeqa
and 
\beq
\Delta F^{(k)}(P_1^{(k)},\dots,P_M^{(k)},J)=-{1 \over \beta x_{k+1}}\ln \int \left[ \prod_{i=1}^M   P_i^{(k)}dP_i^{(k-1)}\right]  e^{-\beta x_{k+1}\Delta F^{(k-1)} (P_1^{(k-1)},\dots,P_M^{(k-1)},J) }
\label{defDFnuApp}
\eeq
The iterative definition has to be supplemented with the $k=0$ case that reads:
\beqa
\tilde{P}^{(0)}(P_1^{(0)},\dots,P_M^{(0)},J) & \equiv & {1 \over e^{-\beta x_{1}\Delta F^{(0)} (P_1^{(0)},\dots,P_M^{(0)},J) }}\int \left[ \prod_{i=1}^M   P_i^{(0)}du_i\right] \left(   \frac{4 \cosh \beta J \cosh \beta \sum_i u_i}{\prod_{i=1}^M 2\cosh \beta u_i}  \right)^{x_{1}}\times
\non
\\
& \times & \delta\left(u- \tilde{u}\left(\sum_i u_i,J\right) \right)
\label{defP0nuApp}
\eeqa
and 
\beq
\Delta F^{(0)} (P_1^{(0)},\dots,P_M^{(0)},J) \equiv -{1 \over \beta x_{1}}\ln \int \left[ \prod_{i=1}^M   P_i^{(0)}du_i \right] \left(   \frac{4 \cosh \beta J \cosh \beta \sum_i u_i}{\prod_{i=1}^M 2\cosh \beta u_i}  \right)^{x_{1}}
\label{defN0nuApp}
\eeq
where we used the definition \cite{MP1}:
\beq
\tilde{u}(h,J)={1 \over \beta} \arctanh [\tanh \beta J \tanh \beta h]
\eeq
We recall also the relationship between the populations and $\rho(\sigma)$:
\beq
\rho_{P^{(k)}}(\s )=\int P^{(k)}dP^{(k-1)}\prod_{{\cal C}=1}^{x_{k+1}/x_{k}}\rho_{P^{(k-1)}}(\{\s\}_{\cal C} )
\label{itdefApp}
\eeq
In the following we will prove the equivalence between eq. (\ref{varr}) and eq. (\ref{varP}).
We basic step is to prove that the following fundamental equation holds at any level $k$:
\beq
\left[\Tr_{\tau_{\cal C}}\left(\prod_{i=1}^M  \rho_{P_i^{(k)}}(\tau_{\cal C}) \right)\exp \beta J \s_{\cal C}\tau_{\cal C} \right]=e^{-\beta x_{k+1} \Delta F^{(k)}(P_1^{(k)},\dots,P_M^{(k)},J)}\rho_{\tilde{P}^{(k)}}(\s_{\cal C})
\label{assump}
\eeq
In the above equations $\s_{\cal C}$ and $\tau_{\cal C}$ are two sets of $x_{k+1}$ spins and $\tau_{\cal C}\s_{\cal C}=\sum_{a=1}^{x_{k+1}}\s_a \tau_a$.
The proof is iterative: assuming that the equation is satisfied at level $k-1$ we will show that it is also satisfied at level $k$.
In order to do that we divide the $x_{k+1}$ spins $\s_{\cal C}$ in $x_{k+1}/x_k$ groups $\s_{\cal C'}$ of $x_k$ spins and we use the definition (\ref{itdefApp}):
\beqa
\Tr_{\tau_{\cal C}}\left(\prod_{i=1}^M  \rho_{P_i^{(k)}}(\tau_{\cal C}) \right)\exp \beta J \s_{\cal C}\tau_{\cal C}  & = & \Tr_{\tau_{\cal C}}\left(\prod_{i=1}^M  \int P_i^{(k)}dP_i^{(k-1)}\prod_{{\cal C'}=1}^{x_{k+1}/x_{k}}\rho_{P_i^{(k-1)}}(\tau_{\cal C'} ) \right)\exp \beta J \s_{\cal C}\tau_{\cal C} =
\non
\\
& = & \int\left[ \prod_{i=1}^M   P_i^{(k)}dP_i^{(k-1)}\right]\prod_{{\cal C'}=1}^{x_{k+1}/x_{k}}\Tr_{\tau_{\cal C'}}\left[\left(\prod_{i=1}^M  \rho_{P_i^{(k-1)}}(\tau_{\cal C'} ) \right)\exp \beta J \s_{\cal C'}\tau_{\cal C'} \right] 
\label{itint}
\eeqa
Now assuming that eq. (\ref{assump}) holds true at level $k-1$ and integrating over a delta function $\delta(P^{(k-1)}-\tilde{P}^{(k-1)})$ we get:
\beq
\Tr_{\tau_{\cal C}}\left(\prod_{i=1}^M  \rho_{P_i^{(k)}}(\tau_{\cal C}) \right)\exp \beta J \s_{\cal C}\tau_{\cal C} =
\eeq
\beq
=\int dP^{(k-1)} \left\{ \left[ \prod_{i=1}^M   P_i^{(k)}dP_i^{(k-1)}\right]\delta(P^{(k-1)}-\tilde{P}^{(k-1)}) e^{-\beta x_{k+1}\Delta F^{(k-1)} (P_1^{(k-1)},\dots,P_M^{(k-1)},J) }\right\} \prod_{{\cal C'}=1}^{x_{k+1}/x_{k}}\rho_{P^{(k-1)}}(\s_{\cal C'}) 
\eeq
we see that the term in curly brackets corresponds to the one in the definition (\ref{defPnuApp}), and using the definition (\ref{itdefApp}) we conclude that eq. (\ref{assump}) holds true at level $k$. 

In order to complete the proof we need to show that eq. (\ref{assump}) holds for $k=0$.
In this case $\s_{\cal C}$ is a group of $x_1$ spins, using eq. (\ref{itint}) we have:
\beqa
\Tr_{\tau_{\cal C}}\left(\prod_{i=1}^M  \rho_{P_i^{(0)}}(\tau_{\cal C}) \right)\exp \beta J \s_{\cal C}\tau_{\cal C}  & =  & \int\left[ \prod_{i=1}^M   P_i^{(0)}du_i\right]\prod_{{a}=1}^{x_{1}}\sum_{\tau_a}\left[\left(\prod_{i=1}^M \frac{\exp \beta u_i \tau_a}{2 \cosh \beta u_i} \right)\exp \beta J \s_a\tau_a \right] 
\non
\eeqa
now summing over each $\tau_a$ and introducing a delta function $\delta(u-\tilde{u}(\sum_{i=1}^M u_i,J))$ and using the definitions (\ref{defP0nuApp}) and (\ref{defN0nuApp}) we can see that eq. (\ref{assump}) holds true also at level $k=0$.
The equation (\ref{assump}) can now be used to prove the equivalence between (\ref{varr}) and (\ref{varP}).

\section{The Order-Parameter Equation}
In this appendix we report an order parameter expansion of the variational equation (\ref{eqq}).
Expanding equation (\ref{eqq}) for $q_{abcd}$ in powers of the order parameters we get (see appendices \ref{tracs} and \ref{spintraces}):
\beq
q_{abcd}=\frac{M(M-1)}{1-M \tilde{b}_4}\tilde{b}_2^2(q_{ab}q_{cd}+q_{ac}q_{db}+q_{ad}q_{cb})+O(q^3)
\eeq
Substituting this expression in eq. (\ref{eqq}) for $q_{ab}$ we get at the third order in the order parameter $q_{ab}$:
\beq
0=c_1\, q_{ab}+c_2 \,(q^2)_{ab} + c_{3,1}\, (q^3)_{ab}+ c_{3,2}\, q_{ab} ((q^2)_{bb}+(q^2)_{aa})+c_{3,3}\, q^3_{ab}+c_{3,4}\,q_{ab} \Tr Q^2
\eeq
\beqa
c_1 & = & M \tilde{b}_2-1 
\non
\\
c_2 & = & \tilde{b}_2^2 (M^2-M)
\non
\\
c_{3,1} & = & -\tilde{b}_2^3\frac{(M-1)M (M \tilde{b}_4 +M -2)}{M \tilde{b}_4 -1}
\non
\\
c_{3,2} & = &  \tilde{b}_2^3\frac{(M-1)M (M^2 \tilde{b}_4 +M -2)}{M \tilde{b}_4 -1}
\non
\\
c_{3,3} & = & -\frac{2 \tilde{b}_2^3}{3} \frac{(M-1)M (M(2M-1) \tilde{b}_4 +M -2)}{M \tilde{b}_4 -1}
\non
\\
c_{3,4} & = & -\frac{ \tilde{b}_2^3}{2}  \frac{(M-1)M (M^2 \tilde{b}_4 -1)}{M \tilde{b}_4 -1}
\non
\eeqa
The coefficients of the previous expansion are different from what could be obtained by differentiating the variational expression eq. (\ref{phipow}) derived above. This can be understood noticing that the equation for the order parameter corresponds to the following expression:
\beq
0=\Tr \left[ \s_a \s_b \left( \rho(\s) - \frac{\Tr_{\tau} \rho^M(\tau)\langle \exp J \sum_c \s_c \tau_c \rangle }{\Tr \rho^M(\tau)} \right) \right]
\label{op}
\eeq
while the equation one obtains by differentiating eq. (\ref{PHIdil}) corresponds to:
\beq
0=\Tr \left[ \rho^{M-1}(\{\s\}) \s_a \s_b \left( \rho(\s) - \frac{\Tr_{\tau} \rho^M(\tau)\langle \exp J \sum_c \s_c \tau_c \rangle }{\Tr \rho^M(\tau)} \right) \right]
\eeq
Thus the two expressions are equivalent in the sense that they have the same solution at the order at which they are valid.
It can be checked explicitly that the coefficient $c_{3,4}$ (as much as $a_{4,4}$) vanishes at zero-th order in the expansion in $\tau$, noticing that at $T=T_c$ we have $\tilde{b}_2=1/M$ and $\tilde{b}_4=1/M^2$ (because $J =\pm \tilde{J}$).

In the Sherrington-Kirkpatrick limit $M \rightarrow \infty$ and $J=\tilde(J)/\sqrt{M}$ with  $\overline{J^2}=1$ the coefficients of the order parameter equation go to the corresponding SK limit as can be also seen noticing that in this limit eq. (\ref{op}) reduces to the corresponding SK equation:
\beq
q_{ab}=\frac{\Tr \s_a \s_b \exp [\beta^2 \sum_{a<b}Q_{ab}\s_a\s_b]}{\Tr \exp [\beta^2 \sum_{a<b}Q_{ab}\s_a\s_b]}
\eeq

\section{Traces of $\rho(\s)$}
\label{tracs}
In this section we report various quantities that are relevant to compute the expansions in the order parameter.
We define:
\beq
\rho(\s)=b_0 (1+\tilde{g}(\s))
\eeq 
\beq
\tilde{g}(\s)=\tilde{b}_2\sum_{a<b}q_{ab} \s_a \s_b + \tilde{b}_4\sum_{a<b<c<d}q_{abcd} \s_a \s_b \s_c \s_d + \dots
\eeq 
Then the following traces are needed to compute the expansion of the variational expression of $\Phi(n)$. In order to compute them we need also the traces over spin reported in the next appendix.
\beqa
{1 \over 2^n}\Tr \tilde{g} & = & 0
\\
{1 \over 2^n}\Tr \tilde{g}^2 & = & {\tilde{b}_2^2 \over 2}\Tr q^2+
\tilde{b}_4^2 \left( \frac{M(M-1)}{1-M \tilde{b}_4} \tilde{b}_2^2  \right)^2
\left( {1 \over 8} (\Tr q^2)^2+{1 \over 4} \Tr q^4 -\sum_{abc}q^2_{ab}q^2_{ac}+{1 \over 2} \sum_{ab}q_{ab}^4 \right)
+O(q^5)
\\
{1 \over 2^n}\Tr \tilde{g}^3 & = & \tilde{b}_2^3 \Tr q^3+
3 \tilde{b}_4 \tilde{b}_2^2 \left( \frac{M(M-1)}{1-M \tilde{b}_4} \tilde{b}_2^2  \right)
\left( {1 \over 4} (\Tr q^2)^2+ \Tr q^4 -4\sum_{abc}q^2_{ab}q^2_{ac}+2 \sum_{ab}q_{ab}^4 \right)
+O(q^5)
\\
{1 \over 2^n}\Tr \tilde{g}^4 & = & \tilde{b}_2^4 
\left( {3 \over 4} (\Tr q^2)^2+ 3\Tr q^4 -6\sum_{abc}q^2_{ab}q^2_{ac}+4 \sum_{ab}q_{ab}^4 \right)
+O(q^5)
\eeqa
In order to sum over $q_{abcd}$ in $\Tr \tilde{g}^2$ we used the following identity valid for a general $A_{abcd}$ symmetric with respect to permutations of its indexes
\beq
\sum_{a<b<c<d}A_{abcd}={1 \over 24}\left( \sum_{abcd}A_{abcd}-6\sum_{abc}A_{aabc}+3\sum_{ab}A_{aabb}+8 \sum_{ab}A_{aaab}-6 \sum_{a}A_{aaaa}   \right)
\eeq
The following traces are needed to compute the expansion of the equation for the order parameter:
\beqa
{1 \over 2^n}\Tr \s_a \s_b \tilde{g} & = & \tilde{b}_2 q_{ab}
\\
{1 \over 2^n}\Tr \s_a \s_b \tilde{g}^2 & = & 
2 \tilde{b}_2^2 (q^2)_{ab}+
\non
\\
& + & 2 \tb_2 \tb_4 \left( \frac{M(M-1)}{1-M \tilde{b}_4} \tilde{b}_2^2  \right)
\left(  (q^3)_{ab}-2 q_{ab}((q^2)_{aa}+(q^2)_{bb})+2 q_{ab}^3+{1 \over 2} q_{ab} \Tr q^2  \right)
+O(q^4)
\\
{1 \over 2^n}\Tr \s_a \s_b \tilde{g}^3 &  = & 
\tb_2^3 \left( 6 (q^3)_{ab}-6 q_{ab}((q^2)_{aa}+(q^2)_{bb})+4 q_{ab}^3 +{3 \over 2} q_{ab} \Tr q^2  \right)
+O(q^4)
\eeqa
\beq
b_0^M/\Tr \rho^M = 1- {1 \over 4} M(M-1) \tb_2^2 \Tr q^2+ O(q^3)
\eeq
The next traces are needed to compute the equation for $q_{abcd}$:
\beqa
{1 \over 2^n}\Tr \s_a \s_b \s_c \s_d \tilde{g} & = & \tb_4 q_{abcd}
\\
{1 \over 2^n}\Tr \s_a \s_b \s_c \s_d \tilde{g}^2 & = &
2 \tb_2^2 (q_{ab}q_{cd}+q_{ac}q_{bd}+q_{ad}q_{cb})  
+O(q^3)
\eeqa

\section{Spin Traces}
\label{spintraces}

In the following we report the values of traces over the spins. They have been computed using the following general formula
\beqa
{1 \over 2^n} \Tr \s_a \s_b \s_c \s_d \dots \s_e \s_f \s_g \s_h & = & \sum_{\pi} \delta_{ab}\delta_{cd}\dots \delta_{ef} \delta_{gh}-2\sum_{\pi} \delta_{abcd} \dots \delta_{ef} \delta_{gh}+
\non
\\
& + &16 \sum_{\pi} \delta_{abcdef} \dots \delta_{gh}+4 \sum_{\pi} \delta_{abcd} \delta_{efgh} \dots+\cdots
\label{formula}
\eeqa
The above expression represents the fact that each of the spins $\sigma_a,\sigma_b,\dots$ must appear an even number of times in order for the trace to be non zero. 
The first term describes the case in which each spin appears just two times in the sum and the index $\pi$ runs over all different permutations of the indexes that change $\delta_{ab}\delta_{cd}\dots \delta_{ef} \delta_{gh}$.
The second term describes the case in which one spin appears four times and all the other appear two times. However if this is the case the first term also give a non-zero contribution, for this reason the second term has the factor $-2$ in front of it, because that the l.h.s. of (\ref{formula}) is either $0$ or $1$. Again the index $\pi$ runs over all permutations of the indexex that change the summand. The third term corresponds to the case in which one spin appears six times in the sum, while the fourth corresponds to the case in which two diffent spins appears four times each in the sum.
To give an example, in the case of four spins expression (\ref{formula}) specializes to
\beq
{1 \over 2^n} \Tr \s_a \s_b \s_c \s_d   =  \delta_{ab}\delta_{cd}+\delta_{ac}\delta_{bd}+\delta_{ad}\delta_{cb}-2 \delta_{abcd}
\eeq

Using these expression to couple the replica indexes we get:
\beqa
{1 \over 2^n}\Tr \left( \sum_{ab}q_{ab}\s_a \s_b \right) & = & 0
\\
{1 \over 2^n}\Tr \left( \sum_{ab}q_{ab}\s_a \s_b \right)^2 & = & 2 \Tr q^2
\\
{1 \over 2^n}\Tr \left( \sum_{ab}q_{ab}\s_a \s_b \right)^3 & = & 8 \Tr q^3
\\
{1 \over 2^n}\Tr \left( \sum_{ab}q_{ab}\s_a \s_b \right)^4 & = & 48 \Tr q^4-96 \sum_{abc} q^2_{ab}q^2_{ac}+64 \sum_{ab}q^4_{ab}+12 (\Tr q^2)^2
\eeqa
Other traces necessary to the expansions are:
\beqa
{1 \over 2^n}\Tr \left( \sum_{ab}(q^2)_{ab}\s_a \s_b \right) \left( \sum_{ab}q_{ab}\s_a \s_b \right)^2 & = & 4 \Tr q^4-8 \sum_{abc} q^2_{ab}q^2_{ac}+2 (\Tr q^2)^2
\\
{1 \over 2^n}\Tr \left( \sum_{mn}q_{mn}\s_m \s_n \right)\s_a\s_b & = & 2q_{ab}
\\
{1 \over 2^n}\Tr \left( \sum_{mn}q_{mn}\s_m \s_n \right)^3\s_a\s_b & = &
48 (q^3)_{ab}-48 q_{ab}((q^2)_{aa}+(q^2)_{bb})+32 q_{ab}^3+12 q_{ab} \Tr q^2
\\
{1 \over 2^n}\Tr \left( \sum_{mn}(q^2)_{mn}\s_m \s_n \right) \left( \sum_{mn}q_{mn}\s_m \s_n \right)\s_a \s_b  & = & 8 (q^3)_{ab}+2 q_{ab}\Tr q^2 -4 q_{ab}((q^2)_{aa}+(q^2)_{bb})
\eeqa

\end{document}